\documentclass[prb,twocolumn,showpacs,nofootinbib]{revtex4}
\usepackage[dvips]{graphicx}
\usepackage{color,array,dcolumn}
\usepackage{amsmath}
\usepackage{amssymb}

\begin{document}

\title{Dynamical spin properties of confined Fermi and Bose systems in presence of spin-orbit coupling}

\author{A. Ambrosetti, L. Salasnich, P.L. Silvestrelli}
\email{alberto.ambrosetti@pd.infn.it}
\affiliation{Dipartimento di Fisica e Astronomia ``Galileo Galilei'', University of Padova, via Marzolo 8, I--35131, 
Padova, Italy}

\begin{abstract}
\date{\today}
Due to the recent experimental progress, tunable spin-orbit (SO) interactions represent ideal candidates
for the control of polarization and dynamical spin properties in both quantum wells and cold atomic systems.
A detailed understanding of spin properties in SO coupled systems is thus a compelling prerequisite
for possible novel applications or improvements in the context of spintronics and quantum computers. 
Here we analyze the case of equal Rashba and Dresselhaus couplings in both homogeneous and laterally confined
two-dimensional systems. Starting from the single-particle picture and subsequently introducing two-body
interactions we observe that  periodic spin fluctuations can be induced and maintained 
in the system. Through an analytical derivation we show that the two-body interaction does not involve 
decoherence effects in the bosonic dimer,
and, in the repulsive homogeneous Fermi gas it may be even exploited in combination with the SO coupling to
induce and tune standing currents.
By further studying the effects of a harmonic lateral confinement --a particularly interesting case for Bose condensates-- 
we evidence the possible appearance of non-trivial {\it spin textures}, whereas the further application of
a small Zeeman-type interaction can be exploited to fine-tune the system polarizability.
\end{abstract}

\maketitle

\section{Introduction}
Due to the growing interest in the fields of spintronics and quantum computation, spin manipulation techniques have 
witnessed considerable scientific interest \cite{cond2d,cond3d,condprl,macdonald,randeria,Ambrogas,ambrowell,valin,ci2,Pededress}, 
and have permeated different areas of solid state physics.
While spin control is conventionally accomplished through the application of magnetic fields, spin-orbit (SO) couplings are currently 
emerging as a promising alternative \cite{Mei,Rashcurr}, where the intrinsic momentum dependence of SO effects can be exploited in 
order to modulate non-local magnetization properties and spin transport. 
In particular, the Rashba \cite{Rashba} and Dresselhaus \cite{dress} SO interactions, which couple the particle momentum to spin, 
have emerged in the last years in the context of semiconductor quantum wells, due to their peculiar strength 
tunability \cite{nitta,engels,nittathick}.
Recently, combined Rashba and Dresselhaus SO couplings have also been realized in two-component cold atom systems through the 
application of controlled laser beams \cite{Dalibard,cold,so2,so3}, opening the way to a cleaner control of polarization effects.
The tunability of two-body interaction \cite{Feshbach}, of SO coupling strength and of confining potentials, 
combined with the absence of phononic vibrations, in fact, make cold atom systems ideal candidates 
for investigating static and dynamical polarization effects.

Depending on the specific system and on the details of SO coupling, different situations may be encountered both 
in Bose \cite{bc1,bc2,bc3,bc4,bc5,bc6,bc7,bc8,bc9} and Fermi system \cite{bec1,bec2,bec3,bec4,bec5,bec6,bec7,bec8,bec9,bec10,bec11,bec12,bec13,bec14,bec15,bec16,bec17,bec18,bec19,bec21,Tempere,ambrovmc}.
Concerning the interplay of SO coupling and spin orientation,  a pure Rashba
coupling was shown for instance to hinder the onset of finite polarization (Stoner instability) in both the two- (2D) and three-dimensional (3D) repulsive Fermi gas \cite{ambrovar,ambrodmc}, determining at the same time a more gradual transition to the fully 
polarized state.
A comprehensive understanding of the role of SO couplings in polarization and dynamical spin properties of atomic systems,
however, is still missing, and appears at the moment highly non-trivial due to the difficulties related to treating non-local 
interactions and the contextual emergence of different spin channels, such as singlets and triplets.

In order to tackle this problem, we will consider here the case of low-dimensional confined  gases (that are mostly interesting
in the context of transport), in presence of SO coupling. 
Two-component systems --labeled by a pseudospin index-- will be taken into consideration.
Given the potentially countless SO realizations, we will focus 
in particular on equal Rashba and Dresselhaus interactions, as experimentally realized by Dalibard and coworkers \cite{Dalibard}.
This type of SO coupling admits wave vector independent single-particle spinors, and it is therefore especially relevant for the
realization and control of polarized states.
With the aim of establishing the combined role of dimensionality and confinement, we will first investigate the case of a 2D
 gas, commenting on the role of Fermi and Bose-Einstein statistics, and considering the effect of time propagation on polarization. 
The role of a contact two-body interaction will be also analyzed for both Fermi and Bose system, in relation to polarization effects 
and spin dynamics of a two-particle system. 
Moreover, we will show how the combined effect of SO coupling and two-body repulsion can be exploited in order to induce a 
controllable current in the atomic Fermi gas.
The role of an external harmonic confinement will be finally assessed, in order to evidence differences and analogies
with the homogeneous case, both in static and dynamical context. The derived analytical single-particle ground state solutions will
have a particular relevance in the description of trapped Bose condensates, where coherent fluctuating polarization
and non-trivial spin texture may be encountered.

\begin{figure*}[ht]
\vspace{0.7cm}
\centering
\includegraphics[scale=0.4]{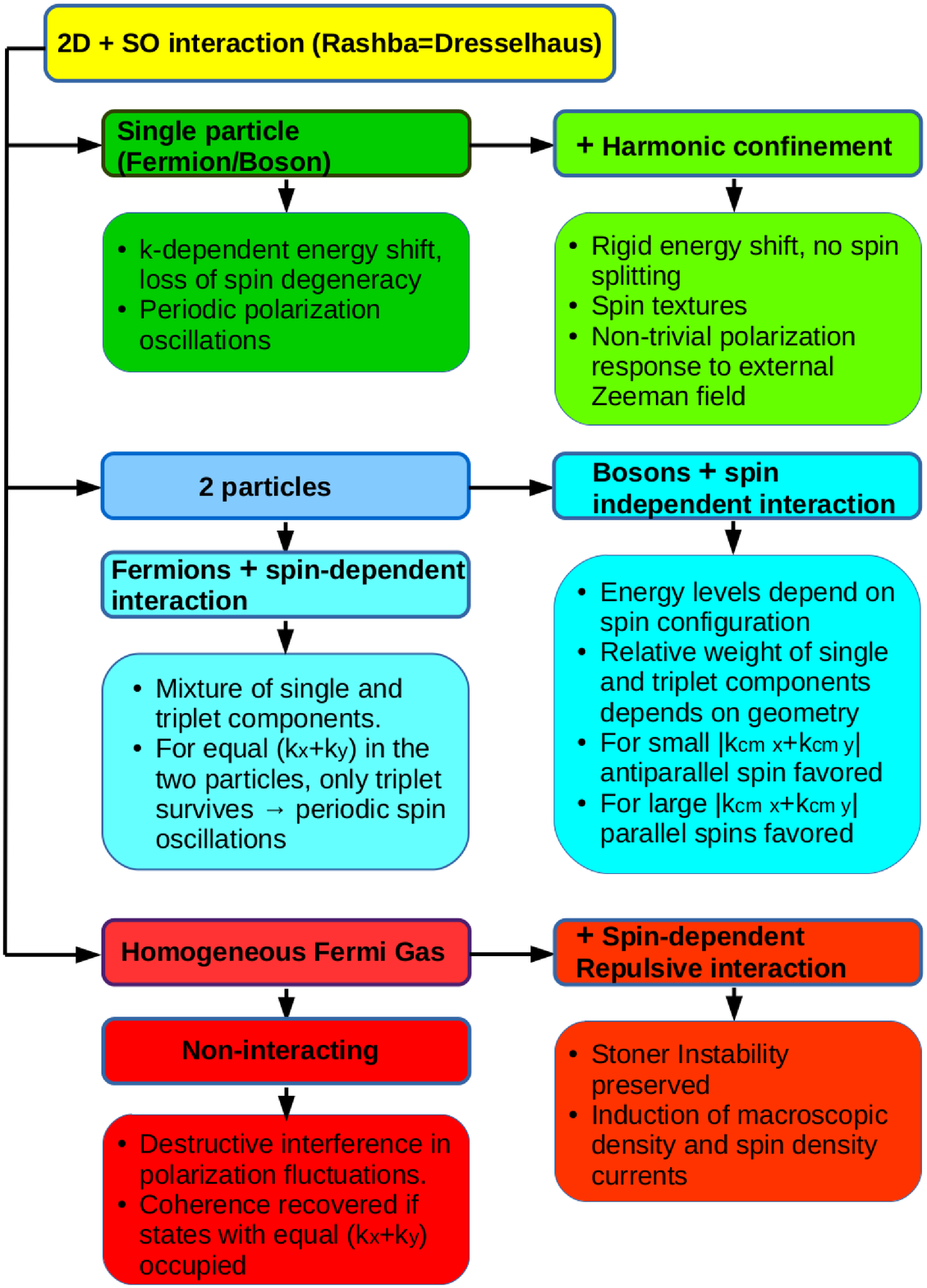}
\vspace{0.5cm}
\caption{(color online) Summary of the considered systems and corresponding results.}
\label{figura1}
\end{figure*}

\section{Spin-Orbit coupling}
\label{sec2}
As mentioned above, the present work focuses on a particular combination of the Rashba and Dresselhaus SO couplings.
Within a second quantization approach, we define the two-components operators 
\begin{equation}
\hat{\Psi}(\mathbf{r})=  \left(
    \begin{array}{c}
      \hat{\psi_{\uparrow}}(\mathbf{r})  \\
      \hat{\psi_{\downarrow}}(\mathbf{r})
    \end{array} \right) 
\qquad  {\rm and} \qquad
\hat{\Psi}^{\dagger}(\mathbf{r})=\left(\hat{\psi}^{\dagger}_{\uparrow}(\mathbf{r}),\hat{\psi}^{\dagger}_{\downarrow}(\mathbf{r})  \right)
\end{equation}
where $\hat{\psi}^{\dagger}(\mathbf{r})_{\sigma}$ and $\hat{\psi}(\mathbf{r})_{\sigma}$ 
construct and annihilate  one particle (fermion or boson) at position $\mathbf{r}$ with (pseudo-)spin $\sigma$ ($\sigma=\uparrow,\downarrow$) respectively.
More precisely, the field operators can be expressed as
\begin{eqnarray}
\hat{\psi}(\mathbf{r})_{\sigma}=\sum_i \phi_i^{\sigma}(\mathbf{r}) \hat{a}_i^{\sigma} \nonumber \,, \\
\hat{\psi}(\mathbf{r})_{\sigma}^{\dagger}=\sum_i \phi_i^{\sigma *}(\mathbf{r}) \hat{a}_i^{\sigma \dagger} \,,
\end{eqnarray}
where $\phi_i^{\sigma}$ for $i\in\{1,\infty\}$ represents a suitable single-particle basis set ({\it i.e.} given by the solution of the one-body Hamiltonian), and the operator $\hat{a}_i^{\sigma}$ ($\hat{a}_i^{\sigma \dagger}$)
annihilates (creates) a particle in the corresponding state.
In order to simplify the notation, the constant $\hbar$ will be set to 1 throughout the paper.
The one-body energy terms in the Hamiltonian will thus be written as
\begin{equation}
\hat{H}_{\rm{1B}}=\int d^2 r\, \hat{\Psi}^{\dagger}(\mathbf{r}) \hat{h}_{sp} \hat{\Psi}(\mathbf{r})\,. 
\end{equation}
where, in the present case,  $\hat{h}_{\rm{sp}}$ contains kinetic energy $\hat{p}^2/(2m)$ and SO interactions.
Following from the above notation, the two SO couplings read:
\begin{eqnarray}
\hat{v}_{{\rm Rash}}=\alpha(\hat{p}_y\sigma_x-\hat{p}_x\sigma_y) \,,\nonumber \\
\hat{v}_{{\rm Dress}}=\beta(\hat{p}_x\sigma_x-\hat{p}_y\sigma_y) \,,
\end{eqnarray}
where $\hat{p}_{x,y}$ represents the momentum operator and $\sigma_{x,y}$ are Pauli matrices.
The factors $\alpha$ and $\beta$ account for the tunability of the SO strengths, and are assumed to be equal in the following.
The overall SO coupling, at $\alpha=\beta$ can thus be expressed as 
\begin{equation}
\label{vso}
\hat{v}_{{\rm SO}}=\alpha(\hat{p}_x+\hat{p}_y)(\sigma_x-\sigma_y) \,.
\end{equation}
Despite the non-locality of the SO coupling, which derives from  momentum dependence, it is clear that $V_{{\rm SO}}$
can be diagonalized in spin space along a single direction, independently from the operator $\hat{p}$. 
This property is peculiar of equal Rashba and Dresselhaus couplings, and does not hold for instance in the case of
pure Rashba or pure Dresselhaus coupling.
Making use of $\hat{z}$ spin components
$\uparrow$ and $\downarrow$ ($\sigma_z$ eigenvectors corresponding to the eigenvalues $+1$ and $-1$), 
the spin eigenstates can be expressed as 
\begin{equation}
\label{spinorpm}
\chi_{\pm}=\left(\pm\frac{(1+ i)}{\sqrt{2}}|\uparrow\rangle+|\downarrow\rangle\right)/\sqrt(2) \,.
\end{equation}
To simplify the notation one can also define the following spin matrix:
\begin{equation}
\label{sigma-}
\tilde{\sigma}_-=(\sigma_x-\sigma_y)/\sqrt{2}\,. 
\end{equation}
The diagonalization of $\hat{v}_{{\rm SO}}$ then follows, observing that $\tilde{\sigma}_-\chi_{\pm}=\pm \chi_{\pm}$.

\section{2D gas}
We consider here the case of a homogeneous 2D gas (either fermions or bosons). We will initially focus on
the single particle problem to understand the spin structure of the system, introducing
two body interactions in a second step.

\subsection{Single-particle problem}
\label{singlehomogeneous}
 In this case, the one-body Hamiltonian contains only
kinetic energy and SO interaction:
\begin{equation}
\hat{h}_{sp}=\frac{\hat{p}^2}{2m}{\mathbb I}+\hat{v}_{{\rm SO}}\,,
\end{equation}
where ${\mathbb I}$ is the 2D identity matrix.
The single-particle spinors which diagonalize the Hamiltonian are 
\begin{equation}
\label{eigenstates}
\phi^{\pm}_{\mathbf{k}}(\mathbf{r})=e^{i\mathbf{k}\cdot\mathbf{r}}\chi_{\pm} \,,
\end{equation}
where $\mathbf{k}$ is the wave vector, and the corresponding eigenenergies are
\begin{equation}
\label{bandso}
\epsilon_{\mathbf{k},\pm}=\frac{\mathbf{k}^2}{2m}\pm \sqrt{2}\alpha (k_x+k_y)\,.
\end{equation}
The introduction of the SO interaction with equal Rashba and Dresselhaus strengths thus induces
a spin degeneracy removal, with  two single-particle parabolic bands shifted in momentum by $\mp 2 \alpha m$
along the direction $(\hat{x}+\hat{y})/\sqrt{2}$.
\begin{figure}[ht]
\vspace{0.7cm}
\centering
\includegraphics[scale=0.35]{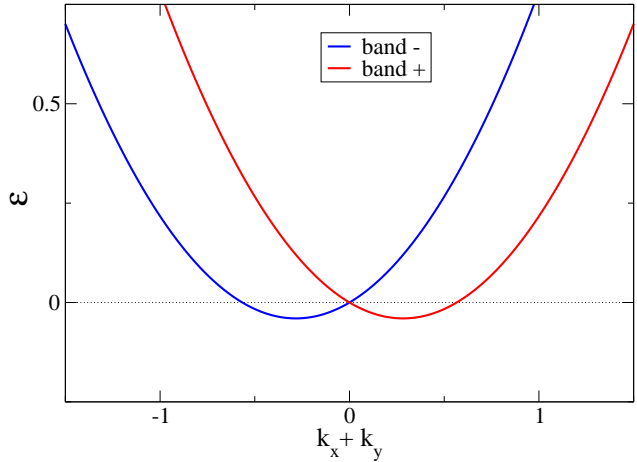}
\vspace{0.5cm}
\caption{(color online) Single-particle bands, as from Eq.\eqref{bandso} computed at $(k_x+k_y)=0$. Lengths are given in units of $n^{-1/2}$, and energies in units of $n/m$. The SO coupling constant $\alpha$ is fixed to 0.2 consistently with these units.}
\label{figura1}
\end{figure}
Following from section \ref{sec2}, all single-particle states relative to the above bands are polarized along
the same direction, at variance with pure Rashba or Dresselhaus eigenstates.
Hence, when defining the single-band occupations as the expectation values of
\begin{equation}
\label{polariz}
\hat{n}_{\pm}=\frac{1}{V}\int d^2r \, \hat{\Psi}^{\dagger}(\mathbf{r}) \frac{\left( {\mathbb I} \pm\tilde{\sigma}_- \right)}{2} \hat{\Psi}(\mathbf{r})\,
%\hat{n}_{\pm}=\frac{1}{V\sqrt{2}}\int d^2r \, \hat{\Psi}^{\dagger}(\mathbf{r}) \left( \sigma_x-\sigma_y \right) \hat{\Psi}(\mathbf{r})\,
\end{equation}
($V$ is the {\it volume} of the system, and $n=n_++n_-$ the total density) 
equal occupancy of $+$ and $-$ bands would lead to zero overall polarization, while
unequal occupancy will cause spin polarization along the direction $(\hat{x}-\hat{y})$, as follows from 
\eqref{vso}.
In analogy with Eq.\eqref{polariz}, one could define the quantities $n_{\uparrow}$ and $n_{\downarrow}$ substituting 
the matrix $\tilde{\sigma}_-$ with $\sigma_z$. Clearly, in this case $n_{\uparrow}-n_{\downarrow}$
corresponds to the spin polarization of the system along $z$.

\subsection{Single-particle wave function evolution}
\label{evolutionsingle}
Making use of the  eigenstates in Eq.\eqref{eigenstates}, and of the corresponding eigenenergies of
Eq.\eqref{bandso}, the time evolution of single-particle wavefunctions can be easily derived. In particular,
we consider here the interesting case of a particle (fermioni or boson) initially in the spin state $\uparrow$, with momentum $\mathbf{k}$,
which evolves after the SO coupling is switched on. 
The initial wave function (at time $t=0$) can be written in spinorial form as
\begin{equation}
\phi(\mathbf{r},t=0)=\frac{1}{\sqrt{V}}\chi_{\uparrow}e^{i\mathbf{k}\cdot\mathbf{r}}
\end{equation}
By making use of Eq.\eqref{spinorpm} one has
\begin{eqnarray}
\phi(r,t)=\frac{1}{1+i} e^{i\mathbf{k}\cdot\mathbf{r}} e^{-i\frac{t k^2}{2m}} \big( e^{-i\sqrt{2}\alpha(k_x+k_y)t}\chi_+ \nonumber \\
-e^{+i\sqrt{2}\alpha(k_x+k_y)t}\chi_-\big) \,, \,
\end{eqnarray}
which can be also recast in the following form
\begin{eqnarray}
\label{spinrot}
\phi(r,t)=\frac{1}{V} e^{i\mathbf{k}\cdot\mathbf{r}}e^{-it\frac{k^2}{2m}} \big( \cos(\sqrt{2}\alpha(k_x+k_y)t)\chi_{\uparrow}-  \nonumber \\
-\frac{\sqrt{2}i}{1+i}\sin(\sqrt{2}\alpha(k_x+k_y)t)\chi_{\downarrow}\big) \,. \qquad
\end{eqnarray}

The $z$ polarization of the system $P_z=n_{\uparrow}-n_{\downarrow}=$ can thus be computed at the generic time $t$, yielding
\begin{equation}
\label{pz}
P_z(t)=\cos(2\sqrt{2}\alpha(k_x+k_y)t)\,.
\end{equation}
According to the above analysis the polarization is homogeneous in space, and oscillates with period $\pi/(\sqrt{2}\alpha(k_x+k_y))$. 

Considering now a homogeneous gas of non-interacting fermions, one finds that
an initially unpolarized system can only evolve in states characterized by $P_z(t)=0$. This can be easily understood given that
the $z$ polarization of a single-particle state with spinor $\chi_{\downarrow}$ at $t=0$ is the opposite of Eq.\eqref{pz}.
As a result, the contributions relative to $\chi_{\uparrow}$ and  $\chi_{\downarrow}$ exactly cancel out for every wave 
vector $\mathbf{k}$.
In contrast, by considering a polarized initial state with only spin $\uparrow$ particles,  the time-dependent polarization becomes
\begin{equation}
\label{pz2}
P_z(t)=\frac{1}{(2\pi)^2}\int d^2 k\, \theta(k-k_F)\cos(2\sqrt{2}\alpha(k_x+k_y)t)\,,
\end{equation}
where $k_F$ is the initial Fermi momentum of the system.
From this formula it is clear how polarization contributions relative to different wave vectors $\mathbf{k}$ generally induce 
destructive interference. Yet, we remark that if the system 
is initially prepared in states with equal $(k_x+k_y)$, no destructive interference occurs, and the system polarizability
fluctuates periodically, and homogeneously in space. Clearly, due to the {\it one-dimensionality} of the wave vector subspace satisfying
this condition, the coherence of spin fluctuations in realistic systems (having finite density) will be limited in time.
Small deviations $\Delta k$ from a fixed value of $(k_x+k_y)$ would however result in small frequency bandwidth $\Delta \nu\sim \alpha \Delta k$ and long-lasting coherence, of the order $\Delta T\sim 2\pi / \Delta \omega$.

In the case of non-interacting Bosons, coherent spin fluctuations may be accomplished through multiple occupation of a single initial 
state with $\mathbf{k}\neq0$. On the other hand, analogous considerations to the Fermi case hold in case of 
small deviations of the momentum from a fixed $(k_x+k_y)$ value.

\subsection{Two-body interaction in Fermi systems and spin current}
We now go beyond the single-particle picture, including a two-body interaction $\hat{H}_{\rm{2B}}$ in the Hamiltonian:
\begin{equation}
\hat{H}=\hat{H}_{\rm{1B}}+\hat{H}_{\rm{2B}} \,.
\end{equation} 
At very low energy only the $s$-wave scattering will be relevant, and the interaction can be modelled through a contact
potential acting between opposite spins. The corresponding two-body contribution to the Hamiltonian thus reads
\begin{equation}
\label{interaz}
\hat{H}_{\rm{2B}}=\int d^2 r\, g \hat{\psi}^{\dagger}_{\uparrow}(\mathbf{r}) \hat{\psi}^{\dagger}_{\downarrow}(\mathbf{r}) 
\hat{\psi}_{\downarrow}(\mathbf{r})\hat{\psi}_{\uparrow}(\mathbf{r})\,.
\end{equation}
The coupling constant $g$ here is positive, and it may be tuned experimentally through the Feshbach resonance 
mechanism \cite{Feshbach,chin}. 

As a first step, we observe that upon introduction of the SO coupling, the $\pm$ single-particle bands \eqref{bandso}
are parabolic and {\it rigidly shifted} with respect to the bands deriving from pure kinetic energy contribution.
Within a mean field approximation \cite{ambrovar}, one could hence write the total energy as a
functional of the single-particle band occupations:
\begin{equation}
\frac{E_{\rm mf}}{V}= \frac{\pi}{m}(n^2_++n^2_-) + g n_+n_- -2m\alpha^2(n_++n_-)\,,
\end{equation}
where $n_{\pm}$ are obtained as the  expectation values of $\hat{n}_{\pm}$ (See Eq.~\eqref{polariz}) over a given non-interacting state.
This expression differs from the standard 2D Stoner model \cite{cond2d,ambrovar} energy functional only by the 
third term on the right, namely the energy shift $2m\alpha^2$ times the total density.  
At constant total density, the system will hence
show the same polarization transition as the standard 2D
Stoner model.
In fact, a minimization of $E_{\rm mf}$ will yield
zero polarization ($P_z = 0$) for $g < 2\pi/m$, where the
kinetic energy increase upon polarization exceeds the {\it polarizing}
effect of the two-body interaction. Full polarization ($P_z = 1$), instead
is obtained for $g > 2\pi/m$, where the two-body repulsion between particles
with opposite spins dominates over the kinetic contribution.
It is important to remark that the occupation of a single band in presence of SO coupling implies non-zero average
momentum in the system, due to the wave vector {\it offset} of  the $\pm$ bands (see Eq.\eqref{bandso}). 
Hence, the polarized state in presence of SO interaction will be characterized by a finite density current.
In fact, the expectation value of the operator
\begin{equation}
\label{curr}
\hat{\mathbf{J}}= \frac{1}{V}\int d^2 r \, \hat{\Psi}^{\dagger}(\mathbf{r})\hat{\mathbf{p}} \hat{\Psi}(\mathbf{r})\,,
\end{equation}
amounts in modulus to $2\alpha m (n_+-n_-)$, indicating a close relation between density current and spin  polarization.

In addition, given that all single-particle states of a given band share the same spin alignment, the system will be
characterized by a finite spin current, as follows from  the operator
\begin{equation}
\label{currso}
\hat{\mathbf{J}}_{s}= \frac{1}{\sqrt{2}V}\int d^2 r \, \hat{\Psi}^{\dagger}(\mathbf{r})\hat{\mathbf{p}} \left( \sigma_x-\sigma_y \right) \hat{\Psi}(\mathbf{r})\,.
\end{equation}
In fact, the expectation value of $\hat{\mathbf{J}}_{s}$ amounts in modulus to $2\alpha m (n_++n_-)$, which is constantly non-zero
for any finite value of the SO coupling, regardless of polarization.

The Feshbach resonance and the SO coupling may thus be viewed as an operational
approach to induce and control density and spin currents in the system. We further remark that the currents' intensities
depend linearly on $\alpha$, and can thus be fine-tuned in combination with the SO coupling strength.

\subsection{Two-fermion system}
In order to investigate the combined effect of two-body interaction and SO coupling on spin dynamics, we study now the case of two fermions. 
In analogy with the single-particle analysis, we initialize the system into a single Slater determinant $|\Phi\rangle$ with plane-wave 
single-particle states characterized by the wave vectors $\mathbf{k}_1$ and $\mathbf{k}_2$, both in the spin state $\uparrow$.
In order to provide a compact operatorial notation, we express field operators as
\begin{equation}
\hat{\psi}(\mathbf{r})_{\sigma}=\sum_{\mathbf{k},\sigma} \hat{a}_{\mathbf{k},\sigma}  \phi_{\mathbf{k},\sigma}  \,,
\end{equation}
where
\begin{equation}
\phi_{\mathbf{k},\sigma} (r)= \frac{1}{\sqrt{V}}\chi_{\sigma}e^{i\mathbf{k}\cdot\mathbf{r}} \,.
\end{equation}
The conjugate fields $\hat{\psi}^{\dagger}(\mathbf{r})_{\sigma}$ are defined accordingly in terms of $a_{\mathbf{k},\sigma}^{\dagger}$.
The operators $a_{\mathbf{k},\sigma}^{\dagger}$, $\hat{a}_{\mathbf{k},\sigma}$ hence create and annihilate plane-wave single particle states
with spin $\sigma$. The wavefunction at $t=0$ can thus be expressed as $|\Phi(0)\rangle=\hat{a}^{\dagger}_{\mathbf{k}_1,\uparrow} 
\hat{a}^{\dagger}_{\mathbf{k}_2,\uparrow}|0\rangle$, where  $|0\rangle$ is the vacuum state, and
the wave function evolution can be described through the equation
\begin{equation}
|\Phi (t)\rangle = e^{-i\hat{H}t} |\Phi (0)\rangle\,.
\end{equation}
Since the Hamiltonian is sum of non-commuting single-particle ($\hat{H}_{\rm{1B}}$) and two-particle ($\hat{H}_{\rm{2B}}$) terms,
we consider the evolution over an infinitesimal time step ($\delta t$), and approximate the real-time propagator through 
Trotter's formula:
\begin{equation}
e^{-i\hat{H}\delta t}=e^{-i\hat{H}_{\rm{1B}}\delta t}e^{-i\hat{H}_{\rm{2B}}\delta t}+ O(\delta t^2)\,.
\end{equation}
It is now obvious how the two-body interaction does not contribute to the first time-evolution step since the two fermions
have parallel spins. Concerning the one-particle propagator, this is diagonal in momentum, but due to the SO coupling it
will produce a spin rotation, according to Eq.\eqref{spinrot}.
The wave function at time $\delta t$ will thus read
\begin{eqnarray}
|\Phi (\delta t)\rangle = \frac{1}{2} \Theta(\mathbf{k}_1,\delta t) \Theta(\mathbf{k}_2,\delta t) \nonumber \\
\bigg( \frac{1+i}{\sqrt{2}} \cos(\sqrt{2}\alpha(k_{1x}+k_{1y})\delta t)\hat{a}^{\dagger}_{\mathbf{k}_1,\uparrow}+ \nonumber \\
+i\sin(\sqrt{2}\alpha(k_{1x}+k_{1y})\delta t)\hat{a}^{\dagger}_{\mathbf{k}_1,\downarrow} \bigg) \nonumber \\
\bigg( \frac{1+i}{\sqrt{2}} \cos(\sqrt{2}\alpha(k_{2x}+k_{2y})\delta t)\hat{a}^{\dagger}_{\mathbf{k}_2,\uparrow}+ \nonumber \\ 
+i\sin(\sqrt{2}\alpha(k_{2x}+k_{2y})\delta t)\hat{a}^{\dagger}_{\mathbf{k}_2,\downarrow} \bigg) |0\rangle \,,
\end{eqnarray}
where $\Theta(\mathbf{k},t)=\exp(-itk^2/2m)$. 
The two-body interaction, at the next steps will then act on the wave function spin-singlet component, and being non-diagonal in
$\mathbf{k}$, it will {\it spread} the wave function in momentum space.
Analogously to non-interacting 2D gas in presence of SO coupling, time propagation will thus lead to a destructive interference
among spin fluctuations.
We notice, however, that if the initial states satisfy the condition $k_{1x}+k_{1y}=k_{2x}+k_{2y}$, only spin-triplet components 
will be present in the wave function at time $\delta t$. In this case, the spin symmetry of the system will thus preserve the periodicity of 
spin oscillations, since these will not be influenced by the two-body interaction. Remarkably, the present condition on momenta 
coincides with that ensuring the periodicity of spin fluctuations in many particle systems. The considerations of Sec. \ref{evolutionsingle}
regarding the evolution of a many particle system hence extend to the case of interacting particles.

\subsection{Two-boson system}
After considering the case of two interacting fermions, we complete our analysis of 2D homogeneous systems studying the
problem of two interacting bosons in presence of SO coupling.
Given the bosonic symmetry of the system,  a two-body interaction different from Eq.\eqref{interaz}, 
will be considered in this case, which is now independent from spin. The new two-body term in the Hamiltonian
reads:
\begin{eqnarray}
\hat{H}'_{\rm{2B}}=\sum_{\sigma_{1,2,3,4}}\int d^2 r_1\, d^2 r_2\, g \hat{\psi}^{\dagger}_{\sigma_1}(\mathbf{r_1}) \hat{\psi}^{\dagger}_{\sigma_2}(\mathbf{r_2}) \nonumber \\
v(|\mathbf{r_1}-\mathbf{r_2}|) \hat{\psi}_{\sigma_3}(\mathbf{r_1})\hat{\psi}_{\sigma_4}(\mathbf{r_2})\,,
\end{eqnarray}
where $\sigma_i=\pm$, and $v(|\mathbf{r_1}-\mathbf{r_2}|)$ describes the details of the spherically symmetric interaction between two particles
at $\mathbf{r_1}$ and $\mathbf{r_2}$.

In the two-particle system one can formally solve Schroedinger's equation by defining center of mass and relative coordinates
\begin{eqnarray}
\label{trasfr}
\hat{\mathbf{r}}_{\rm{cm}}=\frac{\hat{\mathbf{r}}_1+\hat{\mathbf{r}}_2}{2} \nonumber \\
\hat{\mathbf{r}}_{\rm{rel}}=\hat{\mathbf{r}}_1-\hat{\mathbf{r}}_2 \,
\end{eqnarray}
and the corresponding momenta
\begin{eqnarray}
\label{trasfp}
\hat{\mathbf{p}}_{\rm{cm}}=\hat{\mathbf{p}}_1+\hat{\mathbf{p}}_2 \nonumber \\
\hat{\mathbf{p}}_{\rm{rel}}=\frac{\hat{\mathbf{p}}_1-\hat{\mathbf{p}}_2}{2} \,.
\end{eqnarray}
%The newly defined operators satisfy again the commutation relations 
%$[\hat{\mathbf{p}}_{\rm{cm}},\hat{\mathbf{r}}_{\rm{cm}}]=[\hat{\mathbf{p}}_{\rm{rel}},\hat{\mathbf{r}}_{\rm{rel}}]=-i$,
%and $[\hat{\mathbf{p}}_{\rm{rel}},\hat{\mathbf{r}}_{\rm{cm}}]=[\hat{\mathbf{p}}_{\rm{cm}},\hat{\mathbf{r}}_{\rm{rel}}]=0$.
Since the above relations represent a canonical transformation, the relative and center of mass motions are independent 
and can be factorized. Morevoer, the kinetic energy can be expressed in terms of the new operators as:
\begin{equation}
\frac{\hat{\mathbf{p}}_1^2+\hat{\mathbf{p}}_2^2}{2m}=\frac{\hat{\mathbf{p}}_{\rm{cm}}^2+4\hat{\mathbf{p}}_{\rm{rel}}^2}{4m}\,. 
\end{equation}
Concerning the SO coupling, we observe how, depending on the spin configuration of the two particles, this 
will contribute either to the center of mass or relative motion.
In fact, by considering that the spin matrix $\tilde{\sigma}_-$ can be diagonalized through the spinors $\chi_{\pm}$,
in case of two particles with $(++)$ or $(--)$ spin state the SO couplings relative to the two particles will sum up, contributing
to the center of mass motion. In presence of $(+-)$ or $(-+)$ spin configurations instead, the SO couplings will contribute to
the relative motion. Due to the commutativity between center of mass and relative coordinates, the two problems can thus be
separated, and the wave function of the system will be correspondingly factorized into two terms.

Since the two-body interaction only depends on the relative coordinates, the center of mass motion corresponds to that of
a free particle. Hence, for $(++)$ or $(--)$ spin configuration the center of mass is described by the single particle solutions
of Sec.\ref{singlehomogeneous}, while for $(+-)$ and $(-+)$ spin configurations, standard free-particle plane wave solutions apply.

Concerning the relative motion, the case of $(++)$ and $(--)$ spin configurations exactly corresponds to the relative motion in 
absence of SO coupling. In the following we will thus concentrate on the non-trivial case of $(+-)$ or $(-+)$ spin configurations.
Since we are focusing now on relative coordinates, we drop for simplicity the label ``$\rm{rel}$'' from
both $\mathbf{r}_{\rm{rel}}$ and $\mathbf{p}_{\rm{rel}}$.
At this point, we further simplify the notation by defining the new rotated coordinates:
\begin{equation}
\label{trasfr1}
\hat{r}_p=\frac{\hat{x}+\hat{y}}{\sqrt{2}}\,,   \qquad  \hat{r}_m=\frac{\hat{x}-\hat{y}}{\sqrt{2}}\,,
\end{equation}
and the corresponding momentum operators
\begin{equation}
\label{trasfp1}
\hat{p}_p=\frac{\hat{p}_x+\hat{p}_y}{\sqrt{2}}\,, \qquad \hat{p}_m=\frac{\hat{p}_x-\hat{p}_y}{\sqrt{2}}\,,
\end{equation}
again, according to a canonical transformation. In the above definition, $\hat{x}$, $\hat{y}$ and $\hat{p}_x$,
 $\hat{p}_y$ represent the space and momentum operator components along the $O\hat{x}$ and $O\hat{y}$ axes, respectively.
%which satisfy the commutation relations
%\begin{eqnarray}
%[\hat{p}_p,\hat{r}_p]=-i\,, \qquad [\hat{p}_m,\hat{r}_m]=-i\,, \\ \nonumber
%[\hat{p}_p,\hat{r}_m]=0\,,	  \qquad [\hat{p}_m,\hat{r}_p]=0 \,.\qquad
%\label{commut}
%\end{eqnarray}

With the use of the above definitions, and recalling Eq.\eqref{sigma-}, the single-particle Hamiltonian operator for the relative motion reads
\begin{equation}
\hat{h}_{sp}=\frac{\hat{p}_p^2+\hat{p}_m^2}{m}+4 \alpha \hat{p}_p \tilde{\sigma}_-\,,
\end{equation}
where the SO coupling is now expressed in terms of a single momentum operator and a single spin matrix.

Before proceeding we further define the following operator:
\begin{equation}
\label{lz1}
\hat{L}'_z=\hat{r}_p\hat{p}_{m}-\hat{r}_m\hat{p}_{p} \mp 2\alpha m \hat{r}_m  \,.
\end{equation}
Here the sign $+$ applies to the $(+-)$ spin state, and the $-$ sing to $(-+)$.
This operator provides a generalization of the $z$ angular momentum in presence of SO coupling, which commutes both with
the single particle Hamiltonian and with the two-body interaction. As a consequence, the eigenstates of the system can be chosen
to diagonalize $\hat{L}'_z$.
By expressing $\hat{\mathbf{r}}$ in terms of its modulus $r$ and the angle with respect to the axis $O\mathbf{r}_p$,
one finds that the eigenstates $f_l(\mathbf{r})$ of $\hat{h}_{\rm{sp}}$ relative to the eigenvalue $l$ should obey the following equation:
\begin{equation}
-i\frac{\partial f_l(\mathbf{r})}{\partial \theta} = (l \pm 2\alpha m r \sin(\theta)) f_l(\mathbf{r})\,.
\end{equation}
The solutions of the above equations can be expressed as
\begin{equation}
\label{eigenf}
f_l(\mathbf{r})=C(r) e^{i[l\theta\mp 2\alpha m r \cos(\theta)]}\,,
\end{equation}
where $C(r)$ only depends on the relative distance between the two particles, and $l$ can only be an integer number due 
to periodicity.

At this point, we observe that the single particle Hamiltonian corresponding to the relative
coordinates can be rewritten for the considered spin states ($(+-)$ and $(-+)$ respectively) as
\begin{equation}
\hat{h}_{\rm{sp}}=\frac{1}{m} \left(\hat{r}^{-2} \hat{L}_z^{'2}+\hat{r}^{-2}(\hat{r}\cdot
\hat{p}_{\rm{rel}}\pm 2\alpha m \hat{r}_m)^2 -4\alpha^2m^2\right)\,.
\end{equation} 
Making use of the expression \eqref{eigenf}, one obtains the following relation:
\begin{eqnarray}
\hat{h}_{\rm{sp}} C(r) e^{i[l\theta\mp\alpha m r \cos(\theta)]} = \frac{1}{m} \bigg( \frac{C(r)}{r^2}l^2 
-\frac{1}{r^2}(r^2\partial_r^2 C(r) + \nonumber \\
+r\partial_r C(r))-4\alpha^2 m^2 C(r) \bigg) e^{i[l\theta\mp\alpha m r \cos(\theta)]} \,. \qquad
\end{eqnarray}
After simplifying the phase factor, this equation corresponds to that of a system in absence of SO interaction, 
except for the constant energy shift proportional to $\alpha^2 m$.
The energy levels of the system will thus be rigidly shifted due to the SO coupling, while the energy differences between
the different eigenstates will be unaltered, regardless of the details of the potential $v(r)$.

If the potential $v(r)$ in absence of SO coupling is characterized by the spectrum  $E_i$ ($i=1,\infty$ labels both 
{\it discrete} and {\it continuum} eigenvalues), then
the eigenenergies of the SO coupled system for $(+-)$ or $(-+)$ spin configurations will be
\begin{equation}
E_{1,i}(\mathbf{k}_{\rm{cm}})=\frac{k_{\rm{cm}}^2}{4m} +  E_i - 4\alpha^2 m    \,, \,
\end{equation}
where $\mathbf{k}$ accounts for the center of mass momentum.
On the other hand, if the spin configuration is $(++)$ or $(--)$ the energy spectrum becomes
\begin{equation}
E_{2,i}(\mathbf{k}_{\rm{cm}})=\frac{k_{\rm{cm}}^2}{4m} \pm \sqrt{2}\alpha (k_{\rm{cm}, x}+k_{\rm{cm}, y}) +E_i   \,.
\end{equation}
A comparison between the two spectra indicates that the two spin configurations have equal energy if
$k_{\rm{cm} x}+k_{\rm{cm} y}=\pm 2\sqrt{2}\alpha m$ (the $\pm$ sign refers to the $(++)$ and $(--)$ spin configurations), 
i.e. in the minimum of the single-particle bands (see Eq.\eqref{bandso}).
On the other hand, for smaller values of $|k_{\rm{cm} x}+k_{\rm{cm} y}|$  anti-parallel spin configurations 
are favored, while at larger $|k_{\rm{cm} x}+k_{\rm{cm} y}|$ the ground state has parallel spins.

In order to provide a detailed insight in the spin properties of the system we impose now the correct exchange 
symmetry to the wavefunction, distinguishing between the different spin configurations.
In case of parallel spins (either $(++)$ or $(--)$) the exchange of two particles has the only effect of modifying 
the relative coordinate $\mathbf{r}$ into $-\mathbf{r}$. Hence, a symmetric wavefunction can only be 
constructed by imposing that the quantum number $l$ is even (odd in the fermionic case), exactly as in
absence of SO coupling.

Considering now the case of anti-parallel spins, one finds, due to the symmetry of the problem, that the solutions
$f_l(\mathbf{r}) \chi_{+}\chi_{-}$ and $f_l(-\mathbf{r}) \chi_{-}\chi_{+}$ are degenerate in energy.
The center of mass wave function has been neglected here for simplicity since it can be factorized and it is 
symmetric under particle exchange. A wavefunction with correct symmetry can therefore be constructed through a 
combination of the two states, by imposing the condition
\begin{eqnarray}
A f_l(-\mathbf{r}) \chi_{-}\chi_{+} + B f_l(\mathbf{r}) \chi_{+}\chi_{-}= \nonumber \\
=A f_l(\mathbf{r}) \chi_{+}\chi_{-} + B f_l(-\mathbf{r}) \chi_{-}\chi_{+}\,,
\end{eqnarray}
where $A$ and $B$ are complex coefficients satisfying $|A|^2+|B|^2=1$. The particle exchange symmetry is
satisfied if $A=B$. Considering Eq.\eqref{eigenf}, the bosonic {\it relative} eigenfunctions 
thus read
\begin{equation}
\label{eigbos}
\Phi_{\rm{rel}}(\mathbf{r})=\frac{C(r)}{\sqrt{2}} e^{i l \theta}\left(e^{-i2\alpha \cos(\theta)}\chi_{+}\chi_{-}\pm e^{i2\alpha \cos(\theta)}\chi_{-}\chi_{+}\right)\,,
\end{equation}
where the sign $+$ ($-$) is relative to eigenstates with even (odd) $l$: in fact, the transformation $\mathbf{r} \rightarrow -\mathbf{r}$ is equivalent to a modification of the angle $\theta$ by $\pi$.

A simple rewriting of Eq.\eqref{eigbos}, leads to
\begin{eqnarray}
\label{eigbos2}
\Phi_{\rm{rel}}(\mathbf{r})=C(r) e^{i l \theta}\sqrt{2}\bigg(\cos(2\alpha r \cos(\theta))(\chi_{+}\chi_{-}\pm \chi_{-}\chi_{+})- \nonumber \\
-i\sin(2\alpha r \cos(\theta)) (\chi_{+}\chi_{-}\mp \chi_{-}\chi_{+})\bigg)\,, \,
\end{eqnarray}
and an inspection of this formula evidences how the system is characterized by a combination of triplet
and singlet spin states, where the relative weights of the two components depends on the angle $\theta$.
This supports the results of Devreese {\it et al.}, evidencing how, in presence of SO coupling a superfluid state survives even
under application of rather strong magnetic fields due to the existence of spin-triplet wave function components \cite{Tempere}.
Hence, by defining $x'= r \cos(\theta)$, for even (odd) $l$ depending on the orientation of the dimer different situations emerge: at $x'=N\pi/\alpha$ ( with $N$ 
integer) only the triplet (singlet) component is present, while the relative weight of the singlet (triplet) state reaches its 
maximum at $x=(N+1/2)\pi/\alpha$.
We also observe that, given the dependence of the trigonometric functions on $x$, oscillations  will be generally damped at
large $r$ by the wave function decay in case of bound states.

To further investigate the time evolution of spin polarization, we consider a two-particle bosonic state characterized by center of 
mass momentum $\mathbf{k}_{\rm{cm}}$, energy level $i$ (corresponding to odd angular momentum) and initial spin $\uparrow \uparrow$.
We thus express the wave function at time $t$ in terms of the Hamiltonian eigenstates as:
\begin{eqnarray}
\Phi_2(\mathbf{r},t)C(r) e^{-i(\frac{k^2_{\rm{cm}}}{2m}+E_i)t} \frac{1}{2i}\bigg(\chi_{+}\chi_{+} e^{-i\sqrt{2}\alpha (k_x+k_y)t}+ \nonumber \\
\chi_{-}\chi_{-}e^{i\sqrt{2}\alpha (k_x+k_y)t} -(\chi_{+}\chi_{-}+\chi_{-}\chi_{+}) e^{i4\alpha^2 m t} \bigg) \,.
\end{eqnarray}
The time-dependent $z$ polarizability can then be straightforwardly computed, and gives
\begin{equation}
P_z(t)=2\cos(\sqrt{2}\alpha(k_x+k_y)t) \cos(4\alpha^2 m t) \,.
\end{equation}
Even in this case the polarizability fluctuates periodically, and the two body interaction does not induce
destructive interference. We observe that if $k_x+k_y=2\sqrt{2} \alpha m$ then $P_z(t)$ is positive (or zero) at any time,
so that the average polarization of the system in time is $\langle P_z \rangle_t= 2 \pi$.

We finally remark that the present analysis of interacting bosonic dimers could be extended in principle
to Fermionic pairs with spin-independent interaction.

\section{Harmonically confined 2D particles}
In order to investigate the effect of an external confining potential on spin structure and dynamics, we consider 
a 2D system with harmonic confinement. 
Clear analogies exist between the present system and quantum dots \cite{Reimannrev,Reimannlett,diag1,diag2,Kouwenhoven,Steward},
where SO effects have also been recently investigated \cite{ambrodot,governale,cavalli}.

If the confinement is described by a parabolic potential with frequency $\omega$, the single-particle 
Hamiltonian operator can be written as
\begin{equation}
\label{hamt}
\hat{h}_{\rm sp}=\frac{\hat{p}^2}{2m}+\frac{m\omega^2}{2}\hat{r}^2+\hat{v}_{\rm SO} \,.
\end{equation}
We underline at this point that while single-particle solutions exist for the homogeneous 2D gas in presence of
arbitrary combinations of the Rashba and Dresselhaus coupling, the harmonic confinement strongly complicates
the problem due to non commutativity.
Accordingly, use has been made so far of approximated solutions \cite{malet,lipp}, valid in the limit of {\it weak}
SO couplings. Clearly, such approximate solutions imply loss of accuracy and the impossibility of understanding the
limit of large spin-orbit couplings. 
%Going beyond the existing approximations has thus a particular relevance
%in this context, where unraveling subtle spin details is the main goal.
\newline
In the case of equal Rashba and Dresselhaus SO couplings considered here, however, an
exact analytical diagonalization of the single particle Hamiltonian is again possible, being 
the spinorial structure of single-particle orbitals independent from momentum.
\newline
A possible approach to the diagonalization of the single-particle Hamiltonian consists of rewriting $\hat{h}_{\rm sp}$
as $\hat{h}_x+\hat{h}_y$, where
\begin{eqnarray}
\hat{h}_x= \frac{\hat{p_x^2}}{2m}+\frac{m\omega^2}{2}\hat{x}^2 +\sqrt{2}\alpha \hat{p}_x \tilde{\sigma}_- \,, \nonumber \\ 
\hat{h}_y= \frac{\hat{p_y^2}}{2m}+\frac{m\omega^2}{2}\hat{y}^2 +\sqrt{2}\alpha \hat{p}_y \tilde{\sigma}_- \,.
\end{eqnarray}
Since  $\hat{h}_x$ and $\hat{h}_y$ commute, a common set of solutions can be found diagonalizing the two terms independently. 
\newline

\subsection{Solutions}
\label{par2}
As a first step we
notice that the Hamiltonian \eqref{hamt} commutes with $\tilde{\sigma}_-$ (defined in Eq.\eqref{sigma-}), meaning that the two
operators could be diagonalized by the same set of eigenstates. In particular, the eigenstates of $\tilde{\sigma}_-$
corresponding to the eigenvalues $1$ and $-1$ are the spinors of Eq.\eqref{spinorpm}.
Since $\hat{h}_y$ has exactly the same form as $\hat{h}_x$ once $x$ is exchanged with $y$,
we concentrate on $\hat{h}_x$. 
Setting $\hat{p'}_x=\hat{p}_x{\mathbb I}+\sqrt{2}m\alpha\tilde{\sigma}_-$
and observing that $\tilde{\sigma}_-^2={\mathbb I}$ one can write
\begin{equation}
\hat{h}_x= \frac{\hat{p'}_x^2}{2m}+\left( \frac{m\omega^2\hat{x}^2}{2}-m\alpha^2\right) \cdot{\mathbb I}\,,
\label{ham2}
\end{equation}
which admits the following ground state solutions:
\begin{equation}
 \left\{
 \begin{array}{rl}
   f_+(x)=C \cdot e^{-\frac{m\omega}{2}x^2-iKx}  \\
   \\
   f_-(x)=C \cdot e^{-\frac{m\omega}{2}x^2+iKx}  \\
 \end{array} \right.
\label{sol} 
\end{equation}
where 
\begin{equation}
K=\sqrt{2}m\alpha \,,
\end{equation}
 and $C$ is a numerical constant.
Up to now only the problem for $\hat{h}_x$ has been solved. Analogous results, however, hold also
 for $\hat{h}_y$. In order to obtain a solution for $\hat{h}_{\rm sp}$, one can simply
multiply the scalar functions relative to the $x$ and $y$ variables, i.e.
\begin{equation}
 \left\{
 \begin{array}{rl}
   f_+(x,y)=C \cdot e^{-\frac{m\omega}{2}(x^2+y^2)} e^{-iK(x+y)}  \\
   \\
   f_-(x,y)=C \cdot e^{-\frac{m\omega}{2}(x^2+y^2)} e^{iK(x+y)} \\
 \end{array} \right.
\label{solxy}
\end{equation}
These solutions contain a Gaussian overall factor which is due to the harmonic
confinement present in the Hamiltonian, and show at the same time an oscillatory behavior
and a mixing between the up and down spin states (i.e. $\sigma_z$ eigenstates with $s_z=\pm 1$).
This is a clear effect of the SO interaction which couples $\uparrow$ and $\downarrow$ through $\tilde{\sigma}$.
We remark that the two solutions share the same single energy value, meaning 
that the spin-orbit interaction causes no splitting  between different spin states
and therefore no degeneracy removal. On the other hand, the energy spectrum is influenced by the
presence of the spin-orbit interactions by a negative shift proportional to $\alpha^2$.

This result is seemingly at variance with the homogeneous 2D Fermi gas considered above.
In fact, in the homogeneous gas there exist two non degenerate spin states depending on the wave vector 
$\vec{k}$. The energy splitting,  given by $\alpha2\sqrt{2}|k_x+k_y|$ , has a linear dependence on $\alpha$. 
However, the occupation of the single particle quantum dot ground state within the vanishing confinement limit
would correspond to the occupation of the $|\mathbf{k}|=0$ state in the homogeneous case,
which is again spin degenerate, due to the splitting dependence on $|\mathbf{k}|$.

Before concluding, we stress that an alternative approach to this problem can followed by means of transformed variables as defined
in Eqs.\eqref{trasfr1},\eqref{trasfp1}.
With this transformation, the Hamiltonian is recast in the form $\hat{h}_{\rm sp}=\hat{h}_p+\hat{h}_m$, where
\begin{equation}
\hat{h}_p=\frac{\hat{p}_p^2}{2m}+m\omega^2\hat{r}_p^2+\sqrt{2}\alpha \hat{p}_p \tilde{\sigma}_-
\end{equation}
and
\begin{equation}
\hat{h}_m=\frac{\hat{p}_m^2}{2m}+m\omega^2 \hat{r}_m^2.
\end{equation}
From the canonical transformations one has $[\hat{h}_p,\hat{h}_m]=0$ and the two problems can therefore
be solved independently. Furthermore, only $\hat{h}_p$ depends on the SO interaction, meaning that
the motion for the $m$ variables will be that of a standard harmonic oscillator, regardless of the spin. 
The solutions for $\hat{h}_p$ can instead be derived following the procedure described for 
$\hat{h}_x$. From this point of view it is easily understood why the phases of the solutions \eqref{solxy} only show a dependence 
on momentum variables.

%%%%%%%%%%%%%%

When analyzing the symmetries of the system one finds that the operator $\hat{L}_z$ ($z$-component of the angular momentum), 
which commutes with 
the kinetic energy and external harmonic potential,  does not commute with the spin-orbit interactions. 
Hence, the system is not invariant under in-plane rotations, and its eigenstates will not diagonalize the angular momentum operator.
A less obvious symmetry can be determined by defining a modified angular momentum, in analogy with Eq.\eqref{lz1}:
\begin{equation}
\hat{L}'_z=\hat{x}\hat{p}'_y-\hat{y}\hat{p}'_x \,,
\end{equation}
as shown in the Appendix.

\subsection{Spin properties}
\begin{figure}[ht]
\vspace{0.7cm}
\centering
\includegraphics[scale=0.8]{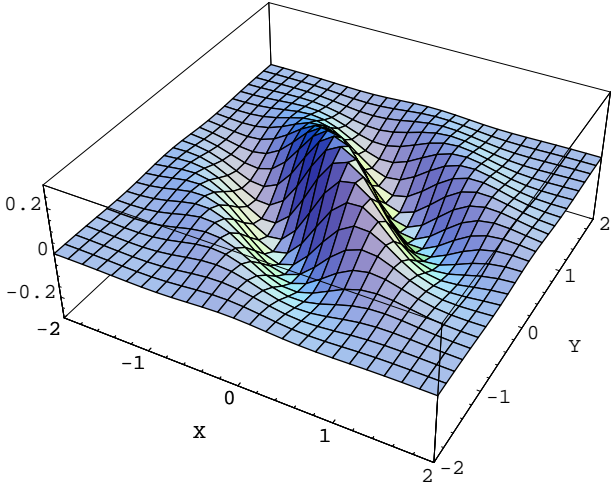}
\vspace{0.5cm}
\caption{(color online) Example of "z spin texture" resulting from the occupation of a single orbital of the type \eqref{solxy} with $c_1=-c_2=1/\sqrt{2}$. 
In the vertical axis $\sigma(\mathbf{r})$ is reported, as defined in Eq. \eqref{spinr}.
Distances are given in units of $1/\sqrt{m\omega}$, and the coupling constant $K$ is set to $1.6 \sqrt{m\omega}$. }
\label{figura}
\end{figure}

Expanding a generic normalized wave function $|\phi\rangle$ over the Hamiltonian eigenstates $|\phi_i\rangle$ as
\begin{equation}
|\phi\rangle=\sum_i c_i |\phi_i\rangle=\sum_i c_i f_i(\mathbf{r})\chi_i
\end{equation} 
one could easily define the expectation value of the operator $\sigma_z$ as
\begin{equation}
P_z=\int d^2 r\, \sigma_z(\mathbf{r}),
\end{equation}
where
\begin{equation}
\label{spinr}
\sigma_z(\mathbf{r})=\sum_{i,j} c_i^* c_j f_i(\mathbf{r})^* f_j(\mathbf{r}) \langle \chi_i|\sigma_z|\chi_j\rangle
\end{equation}
is usually referred to as the $z$ spin density.
\newline
Let us consider first a single particle in the ground state. 
It is easy to show that if the particle is in  $|0^+\rangle$ or  $|0^-\rangle$,
then not only the {\it total polarization} $\langle \phi|\sigma_z|\phi\rangle$ is zero but also $\sigma_z(\mathbf{r})=0$ is,
indicating that that no {\it $z$ spin texture} \cite{ambrodot} (namely space-dependent $\sigma_z(\mathbf{r})$ patterns)
will be present.
This however is not true if the particle occupies any of the states $c_1|0^+\rangle+c_2|0^-\rangle$, with $c_1$ and $c_2$ being 
some numerical constants obeying $|c_1|^2+|c_2|^2=1$.
For example, setting $c_1=-c_2=1/\sqrt{2}$ and calling this state $|\psi_b\rangle$ then one has
\begin{eqnarray}
\sigma_z(\mathbf{r})=\sqrt{\frac{m\omega}{4\pi}} \cos(2K(x+y))e^{-m\omega(x^2+y^2)}
\end{eqnarray}
yielding the "$z$ spin texture" shown in Fig. \ref{figura}. At the same time, integrating this expression over
$\mathbf{r}$ one obtains a non zero net magnetization. The possibility of having both 
presence or absence of spin texture is caused by the ground state being degenerate.

Analogously to the 2D homogeneous case, we also investigate the time evolution of a single particle wave function.
Since the ground state solutions are degenerate in energy, any combination of these would evolve simply acquiring 
an overall phase factor. As a consequence, considering that the $\chi_+$, $\chi_-$ spinors span the entire spin space,
it is hence possible to maintain any arbitrary spin polarization constant in time.
An analogous result can be found in the homogeneous 2D case only through the superposition of single particle states
with momentum $\mathbf{k}$, $\mathbf{k}'$ and spin $+$, $-$ satisfying the property $k_x+k_y=-k'_x-k'_y$.

A different picture, instead emerges when combining eigenstates relative to different energies. For instance, the wave function
$|\phi(0)\rangle=|0^+\rangle + \hat{a}^{'\dagger}_x|0^- \rangle$ (the treatment straightforwardly extends to other excited states) 
would evolve analogously to Eq.\eqref{spinrot}, yielding the following time-dependent polarization density:
\begin{equation}
\sigma_z(\mathbf{r},t)= \sqrt{\frac{2}{\pi}}\,m\omega x \, \cos(2K(x+y)-\omega t) \,.
\end{equation}
This corresponds to a time-dependent spin texture, periodically fluctuating in time, damped at large distance from
the potential well center due to the harmonic confinement. The factor $x$ is due to the specific excited state chosen,
and in the most general case $|\phi(0)\rangle=A \hat{a}^{'\dagger}_x|0^-\rangle + \hat{a}^{'\dagger}_y B|0^-\rangle$ 
(with $|A|^2+|B|^2=1$) 
it will take the form  $(A x+B y)$. The wave-like dependence of the phase on $(x+y)$ is instead an intrinsic property of 
the system, and it is induced by the SO coupling.

In consideration of the previous sections, macroscopic spin textures and even collective polarization fluctuations 
might thus be induced in case of Bose condensates, where the many body wave function allows for a collective occupation of the
ground state.
Besides, we stress that the presence of many fermions in harmonic confinement may also lead to non-trivial spin properties. 
To evidence this aspect we consider the example of two non-interacting particles in the ground state occupying the two spin configuration 
due to antisymmetry. 
This corresponds to a closed shell configuration since all angular momentum
and spin states corresponding to a fixed value of the energy are filled. By explicitly writing
the two occupied states $|0^+\rangle$ and $|0^-\rangle$, the two-particle antisymmetric
wave function describing our system is given by the Slater determinant
\begin{equation}
\phi_{2} = Det\left(
\begin{array}{cc}
   \psi_0^+(\mathbf{r_1,s_1})  & \psi_0^-(\mathbf{r_1,s_1}) \,,  \\
   \psi_0^+(\mathbf{r_2,s_2})  & \psi_0^-(\mathbf{r_2,s_2}) \,,  \\
\end{array} \right) 
\end{equation}
where $\psi_0^+(\mathbf{r,s})$ stands for the spinor corresponding to $|0^+\rangle$ calculated
in $\mathbf{r}$ and contracted to the spin coordinates
$\mathbf{s}=(s^{\uparrow},s^{\downarrow})$.
Also in this case the system shows no magnetization along the $z$ axis and no {\it $z$ spin texture}. 
This is obviously true in general, for the occupation of any of the states \eqref{solxy}.
However, when computing the expectation value of the operator
$\sigma^2=\vec{\sigma}\cdot\vec{\sigma}$ where
$\vec{\sigma}=\vec{\sigma}_1+\vec{\sigma}_2$, one can see how the two particles
do not occupy a spin singlet state: while the expectation value of $\sigma^2$ over
a singlet would be zero, in this case it is not. In fact, one can show that
\begin{eqnarray}
\langle \phi_{2}|\sigma^2|\phi_{2}\rangle= \nonumber \\ \qquad
=\frac{16m^2\omega^2}{\pi^2}\int d^2 r_1 \, d^2 r_2\,
[e^{-m\omega(r_1^2+r_2^2)} \nonumber \\ \qquad
\sin^2(K(x_2+y_2-x_1-y_1))]= \nonumber \\
=8\big(1-exp(-\frac{4K^2}{m\omega})\big)\,.
\end{eqnarray}
This tends to zero for $K\rightarrow 0$ corresponding to the limit of very small SO coupling
and also for $m\omega \rightarrow \infty$ corresponding to overwhelming harmonic confinement.
It is clear from the integral how the contribution comes from an oscillatory behavior damped
by the confinement which gives exponentially decreasing densities.
A deviation of about $10$ percent from singlet $\sigma^2$ expectation value could be obtained 
for instance if the argument of the exponential is about $10^{-1}$. 

\subsection{Linear response}
So far we considered how the spin of the system and its dynamics can be influenced by a SO
coupling. In many cases, however, spins are controlled through external fields, inducing
Zeeman-type interactions.
In this section the polarization induced by a small Zeeman-type interaction will be calculated for
a single-particle system within the static linear response theory\cite{lipp}, in order to unravel the role of 
the SO coupling in the response mechanism. Clearly, some parallelism exists between the procedure followed here and
perturbative approaches for two-level systems\cite{penna}.
In absence of SO coupling the quantum Harmonic oscillator Hamiltonian commutes with $\sigma_z$,
so that an eigenstate of the system initially set for instance in the $\uparrow$ spin state
will be unaltered by the application of the Zeeman interaction. Only additional external perturbations,
or the initial superposition of different eigenstates might lead to a change in the polarization.

In contrast, in the presence of SO coupling, the commutativity of the Hamiltonian with $\sigma_z$
is lost, and the Zeeman interaction may perturb the spin state of the system leading to non-zero 
response.
\newline
The single-particle analytical solutions derived in section \ref{par2} represent the best 
choice for a precise calculation of response
properties, where a correct description of the wave function phase can play a major role.
\newline
In presence of a perturbing Zeeman-type field, the Hamiltonian operator could be written as 
$\hat{h}_{\rm{pert}}=\hat{h}_{\rm{sp}}+\hat{h}_{\rm{int}}$, where the interaction Hamiltonian 
operator is
\begin{equation}
\hat{h}_{\rm{int}}=B \sigma_z.
\end{equation}
where $\hat{h}_{\rm{sp}}$ was defined in \eqref{ham2}, and $B$ is the coupling strength.

Due to the two-fold degeneracy of the ground state, particular attention should be given to
the wave function perturbation induced by the operator $\sigma_z$.
For this reason we define the states
\begin{equation}
|\psi_{1,2}^0\rangle=\frac{|0^+\rangle\pm|0^-\rangle}{\sqrt{2}}\,,
\end{equation}
satisfying the relations $\langle \psi_{1,2}^0 |\sigma_z| \psi_{1,2}^0 \rangle=0$, and
$\langle \psi_{1,2}^0 |\sigma_z| \psi_{2,1}^0 \rangle=\pm\exp{(-K^2/(m\omega))}$.
In practice, these states are chosen in a way to lift the ground state degeneracy at the first perturbative order
in the Zeeman interaction.

Let us now set the initial state of the system to $|\psi_2^0\rangle$ (analogous results can 
be found choosing $|\psi_1^0\rangle$), and rewrite the response function of the system as
\begin{eqnarray}
\label{defresp}
\xi=\lim_{B\rightarrow 0}\frac{\langle \psi_0|\sigma_z|n\rangle\langle n|\hat{h}_{int}|\psi_0\rangle}{B(E_0-E_n)}+ \\ \nonumber
\frac{\langle \psi_0|\hat{h}_{int}|n\rangle\langle n|\sigma_z|\psi_0\rangle}{B(E_0-E_n)} \,,
\end{eqnarray}
where the summation over the states $|n\rangle$ is intended to span over all $\hat{h}_{\rm{sp}}$ eigenstates 
orthogonal to the ground state. $E_0$ and $E_n$ 
respectively denote the $\hat{h}_{\rm{sp}}$ eigenvalues relative to $|\psi_1^0\rangle$ and $|n\rangle$.

In computing the elements of \eqref{defresp} one should notice how 
$\langle \chi_{\pm}|\sigma_z|\chi_{\pm}\rangle$ is identically equal to zero, while terms of the
kind $\langle \chi_{\pm}|\sigma_z|\chi_{\mp}\rangle$ give a non zero contribution.
Besides, one can prove that $\sigma_z$ can excite $|0^{\pm}\rangle$ into higher energy states having orthogonal spin.
at variance with the standard oscillator case in absence of SO.
Denoting $\tilde{a}^{\dagger}_{x,y}|0^{\pm}\rangle$ with $|1_{x,y}^{\pm}\rangle$, it is easily proved
that
\begin{eqnarray}
\langle 0^+|\sigma_z|1_{x,y}^-\rangle=-\sqrt{\frac{2}{m\omega}}iKe^{-\frac{2 K^2}{m\omega}}
\nonumber \\
\langle 0^-|\sigma_z|1_{x,y}^+\rangle=\sqrt{\frac{2}{m\omega}}iKe^{-\frac{2 K^2}{m\omega}}.
\end{eqnarray}
This is caused by the presence of the $\pm iKx$ phases in the Hamiltonian eigenstates \eqref{eighls}
, which cancel out for couples of states with equal spin, but add up when considering
transitions between different spin states, leading to non zero space integrals.
Notice how, for a standard harmonic oscillator, such a phase is absent, forbidding transitions
between ground and excited states.
\newline
In view of the properties just illustrated, one can show that in the expression \eqref{defresp}
the intermediate states $|1_{x,y}^+\rangle$ will give a non zero contribution to the
response function. 
Explicitly computing the transition coefficients to higher energy states, it is
also possible to prove that, for small values of the parameter $\alpha$, other terms in the
summation will only give small contributions to the final result due to a geometrical convergence
controlled by the quantity $2 K^2/(m\omega)$. 
According to the properties just illustrated, the response function to leading order in $K^2/(m\omega)$ is:
\begin{eqnarray}
\xi =
\sum_{i=x,y}\frac{\langle {\psi}_1^0|\sigma_z|{\psi}_{1i}^+\rangle
\langle {\psi}_{1i}^+|\sigma_z|{\psi}_1^0\rangle}{- \omega }+  \nonumber \qquad \\
\frac{\langle {\psi}_1^0|\sigma_z|{\psi}_{1i}^-\rangle
\langle {\psi}_{1i}^-|\sigma_z|{\psi}_1^0\rangle}{- \omega }=  \nonumber \qquad \\
=-\frac{2 K^2}{m \omega^2} e^{-\frac{4 K^2}{m\omega}}\,, \qquad 
\end{eqnarray}
and this interestingly differs from zero.
\newline
In absence of SO interaction, the effect of a Zeeman interaction is simply that of lifting 
the spin degeneracy, without modifying the space dependence of the wave function.
As a consequence, the expectation value of $\sigma_z$ over the interacting wave function
is identical to the non-interacting value, and, following from the definition \eqref{defresp}
the response function will be zero. 
Once the SO interaction is considered, instead, one faces a modification
of the $\sigma_z$ expectation value in presence of Zeeman interaction. Given the negative sign of the
response function, the particle will be forced  to partially align along $-\hat{z}$, or,
equivalently,  the particle will be in a {\it stretched} spin configuration.
\newline
Similar calculations could be carried out for the homogeneous non-interacting Fermi gas.  
In that case it could be shown that $\sigma_z$ only allows for transitions
between states having the same wave vector $\mathbf{k}$. This implies that no {\it spin stretching}
effect will be present under application of Zeeman fields. The finite spin response of the system 
is thus a clear signature  of the combined effects of the SO interaction and the harmonic confinement,
and denotes a {\it gradual} modification of the system polarization.

\section{Conclusions}

We have investigated static and dynamical spin properties of two-dimensional systems
in presence of equal Rashba and Dresselhaus coupling. 
The recent developments in the experimental realization of SO couplings in ultracold atomic systems\cite{galitski,luo},
and the contextual confinement of Bose\cite{yefsa} and Fermi\cite{guan} gases, suggest that the detailed control of spin
effects may become possible in the next years. Theoretical predictions can thus serve as a guide for the next experimental
investigations, suggesting possible pathways for controlling both static and dynamical spin properties.
In the homogeneous gas we
showed how the SO coupling can induce single-particle polarization time fluctuations. 
The oscillations relative to single-particle states with different wave vectors interfere
destructively in general. When occupying states characterized by equal values of $(k_x+k_y)$,
however, collective periodic polarization fluctuations can occur in the system.

Analogous conclusions can be extended to two-particle Fermi systems in presence 
of contact interaction.
In fact, this condition on momenta ensures, for two fermions initially in the $\uparrow$
spin configuration, the absence of spin-singlet wavefunction components, protecting
the wave function from scattering on different momenta.
Periodic polarization oscillations can also take place in interacting two-boson systems,
where an  analytic description of the problem has been accomplished.

Moreover, we have evidenced how density and spin-density currents can be induced and controlled
in the repulsive homogeneous Fermi gas due to the combined effect of SO coupling and
Stoner instability. The experimentally available control of the SO coupling strength and the
Feshbach resonance mechanism hence represent potentially viable tools for an efficient control of the
transport properties of the system.

To complete our study, a 2D system in presence of external harmonic confinement has been taken into account.
Exact single particle solutions have been derived and analyzed, demonstrating the emergence of
complex spin textures, which evolve in time in analogy with the homogeneous case.
These results suggest that Bose condensates, due to {\it macroscopic population} of the ground state may
analogously exhibit non-trivial collective polarization features.
Interestingly, the SO coupling induces a combination of spin singlet and triplet even in a simple
closed-shell two-particle configuration.
Finally, the application of linear response theory indicates that in presence of a SO interaction
a harmonically confined particle can respond to an applied Zeeman interaction, allowing for 
detailed control of the system polarization.

\section{Acknowledgements}
We ackowledge scientific collaboration with Flavio Toigo, who actively contributed to the
present article. We also thank F. Pederiva and E. Lipparini for useful discussion and fruitful suggestions.

\section{Appendix: alternative approach to harmonically confined 2D particles}
Due to the dependence of $\hat{L}'_z$ on $\hat{\mathbf{p}}'$ (which satisfies $[\hat{p}'_x,x]=-i$ and analogously for the $y$ components), 
this operator commutes with the Hamiltonian \eqref{hamt},
so that a common basis set can be defined which contemporarily diagonalizes $\hat{L}'_z$ and $\hat{h}_{\rm{sp}}$. 
Using polar coordinates ($x=r\cos\varphi$ and $y=r\sin\varphi$),
the ground state solutions given in the previous section read 
\begin{eqnarray}
\psi_0^+(r,\varphi) =\left(\frac{m\omega}{4\pi}\right)^{1/4} e^{-iK r(\cos\varphi+\sin\varphi)} e^{-\frac{m\omega}{2}r^2}
\left(
\begin{array}{c}
   \frac{1+i}{\sqrt{2}}  \\
   1  \\
\end{array} \right) \nonumber \\
\psi_0^-(r,\varphi) =\left(\frac{m\omega}{4\pi}\right)^{1/4} e^{iK r(\cos\varphi+\sin\varphi)} e^{-\frac{m\omega}{2}r^2}
\left(
\begin{array}{c}
   -\frac{1+i}{\sqrt{2}}  \\
   1  \\
\end{array} \right) 
\label{eighls}
\end{eqnarray}
It is easily shown that these are at the same time eigenstates of $\hat{h}_{\rm{sp}}$ with energy
 $\omega-2m\alpha^2$, and of $\tilde{L}_z$ with eigenvalue $0$.
Other eigenstates of $\tilde{L}_z$ relative to the eigenvalue $l$ can be obtained
by defining the following creation and destruction operators 
\begin{eqnarray}
\hat{a}_{\pm}^{'\dagger}=\frac{1}{\sqrt{2}}(\hat{a}_x^{'\dagger}\pm i\hat{a}_y^{'\dagger}) \nonumber \\
\hat{a}'_{\pm}=\frac{1}{\sqrt{2}}(\hat{a}'_x \mp i\hat{a}'_y). \nonumber \\
\end{eqnarray}
These operators satisfy the usual commutation properties, in analogy with $\hat{a}^{'\dagger}$, $\hat{a}'$.
 
Moreover, from the application of $\hat{a}_{\pm}^{'\dagger}$ it becomes evident how 
$\hat{L}'_z$ and $\hat{L}_z$ eigenstates only differ by the phase factor $\pm Kr(\cos\varphi+\sin\varphi)$,
induced by the SO coupling.

\end{document}